\documentclass[11pt]{article}
\usepackage{amssymb,amsmath}
\usepackage{epsfig}
\usepackage{pennames}

\begin{document}

\rightline{Ref. SISSA 63/2001/EP}
\rightline{TUM-HEP-427/01}
\rightline{December 2001}
\vskip 0.6cm
\begin{center}
{\bf
CONSTRAINTS ON $|U_{e3}|^{2}$ FROM A THREE-NEUTRINO
OSCILLATION ANALYSIS OF THE CHOOZ DATA}

\vspace{0.3cm}
S. M. Bilenky$^{(a)}$,
\footnote{Also at: Joint Institute for Nuclear Research, Dubna, Russia}
~~D. Nicolo$^{(b,c)}$ and
~~S. T. Petcov$^{(d,e)}$
\footnote{Also at: Institute of Nuclear Research and
Nuclear Energy, Bulgarian Academy of Sciences, 1784 Sofia, Bulgaria}

\vspace{0.3cm}
{\em $^{(a)}$ Physik Department, Technische Universitat Munchen,
James-Franck-Strasse,
D-85748, Garching, Germany \\} 
\vspace{0.2cm}
{\em $^{(b)}$ Dipartimento di Fisica, Universita
di Pisa, I-56127 Pisa, Italy\\
}
\vspace{0.2cm}
{\em $^{(c)}$ Istituto Nazionale di Fisica Nucleare,
Sezione di Pisa, I-56010 Pisa, Italy\\
}

\vspace{0.2cm}
{\em $^{(d)}$ Scuola Internazionale Superiore di Studi Avanzati,
I-34014 Trieste, Italy\\
}
\vspace{0.2cm}
{\em $^{(e)}$ Istituto Nazionale di Fisica Nucleare,
Sezione di Trieste, I-34014 Trieste, Italy\\
}
\end{center}
 
\begin{abstract}
Results of a new analysis of the CHOOZ data,
performed in the framework of three-neutrino mixing, 
are presented. Both the cases of normal and inverted
neutrino mass hierarchy are considered.
The parameters characterizing 
the solar neutrino oscillations,
$\Delta m^2_{\odot}$ and  
$\tan^{2}\theta_{\odot}$, are assumed to
lie in the region of the 
large mixing angle (LMA) 
MSW solution of the
solar neutrino problem,
which is favored by the current
solar neutrino data. 
At $\Delta m^{2}_{\odot} \lesssim 
10^{-4} \rm{eV}^{2}$ 
the new CHOOZ  exclusion curve in
the $\Delta m^{2}_{31} - |U_{e3}|^2$ plane 
practically coincides with 
the exclusion curve  obtained in 
the two-neutrino mixing analysis.
For $\Delta m^{2}_{\odot} \gtrsim 2\cdot
10^{-4}~{\rm eV^2}$, the constraints
of the CHOOZ data on $|U_{e3}|^2$
are more stringent.
For, e.g., $\sin^2\theta_{\odot} = 0.50$
and $\Delta m^{2}_{\odot}= 6\cdot
10^{-4}\,~\rm{eV}^{2}$, and at 
$\Delta m^{2}_{31} = 2.5\cdot 10^{-3}~{\rm eV^2}$
(the Super-Kamiokande best fit-point) 
we find that (at 90\% C.L.)
$|U_{e3}|^{2} < 1.7 \cdot10^{-2}$,
which is by more than a factor of 
2 smaller than the upper bound
obtained in the original 
two-neutrino mixing analysis of the CHOOZ data 
($|U_{e3}|^{2} < 3.7 \cdot10^{-2} $).

\end{abstract}
\newpage
\section{Introduction}
\vskip -0.2truecm
\hskip 0.5truecm There exist at 
present strong evidences  
in favor of neutrino oscillations obtained in 
the atmospheric \cite{SKatm}
and solar neutrino experiments 
\cite{Cl98,Fu96,SAGE,GALLEXGNO,SKsol,SNO1}.
The results of the recent SNO experiment
\cite{SNO1}, combined with the data from 
the Super-Kamiokande experiment \cite{SKsol}, 
which clearly demonstrated the 
presence of $\Pgngm$ ( $\Pgngt$)
in the flux of the solar neutrinos reaching the Earth
\footnote{The non-electron neutrino component 
in the flux of solar neutrinos can also
include, or correspond to, $\Pagngm$ and/or 
$\Pagngt$ \cite{SNO1}.} \cite{SNO1}, 
are additional compelling
evidence in favor of neutrino oscillations. 
All these data suggest the existence of 
mixing of at least three massive neutrinos:
\begin{equation}
\nu_{{\alpha}L} = \sum_{j=1}^{3} U_{\alpha\, j} \, \nu_{jL}, ~~~ 
\alpha = e, \mu, \tau.
\label{001}
\end{equation}
%
\noindent Here $\nu_{{\alpha}L}$
is the field of the flavour neutrino $\nu_{{\alpha}}$,
$\nu_{j}$, $j=1,2,3$, is the field of neutrino with
mass $m_{j}$ and $U$ is a 3x3 unitary  matrix - the 
Pontecorvo-Maki-Nakagawa-Sakata (PMNS) neutrino mixing 
matrix \cite{Pont57,MNS62}. 

 Indications for $\Pagngm \to \bar \Pagne$
oscillations were found in the Los Alamos short baseline
accelerator experiment LSND \cite{LSND}. 
In order to describe the data of the
atmospheric, solar and LSND experiments
in terms of neutrino mixing and oscillations
one needs to assume the
existence of sterile neutrinos and
mixing of (at least) four massive neutrinos
(see, e.g., \cite{BGG99}). The LSND result 
will be checked
by the MiniBooNE experiment \cite{MiniB}
scheduled to start in 2002.
We will consider
in what follows only the case of  
three-neutrino mixing and oscillations.

   The analyses of the Super-Kamiokande 
atmospheric neutrino data suggest that the atmospheric 
$\Pgngm$ ($\Pagngm$) oscillate predominantly
into $\Pgngt$ ($\Pagngt$)   
and that these oscillations are induced by a 
neutrino mass squared difference
$\Delta m^2_{atm}$ having a value in the range
\cite{SKatm} (99\% C.L.)
\begin{equation}
1.3\cdot 10^{-3}~{\rm eV^2} \lesssim \Delta m^2_{atm} \lesssim 
5.0\cdot 10^{-3}~{\rm eV^2}.
\label{dmatm}
\end{equation}
%
\indent   Global analyses of the solar neutrino data 
including the SNO results \cite{SNO1}
were performed, e.g., in 
\cite{FogliSNO,ConchaSNO,GoswaSNO,StrumiaSNO,GiuntiSNO}.
The data were shown to favor the large mixing
angle (LMA) MSW, the LOW and the quasi-vacuum oscillation 
(QVO) solutions of the solar neutrino problem.
In the case of the LMA solution,
the range of values of the corresponding 
neutrino mass-squared difference 
was found 
in \cite{FogliSNO} and \cite{ConchaSNO}
to extend (at 99\% C.L.) up to
values  of $\Delta m^2_{\odot}$ as large as 
$\sim 5.0 \cdot 10^{-4}~{\rm eV^2}$
and $\sim 8.0 \cdot 10^{-4}~{\rm eV^2}$, 
respectively: 
\begin{equation}
{\rm LMA~~MSW~\cite{FogliSNO,ConchaSNO}}:~~~~~~2.0\cdot 10^{-5}~{\rm eV^2} 
\lesssim \Delta m^2_{\odot} 
\lesssim  
(5.0 - 8.0)\cdot 10^{-4}~{\rm eV^2}~. 
\label{dmsol}
\end{equation}
%
\indent The most stringent constraints on the oscillations
of electron (anti-)neutrinos 
were obtained in the CHOOZ
\cite{CHOOZ2} and Palo Verde \cite{PaloV} 
long baseline disappearance 
experiments with reactor $\Pagne$. 
These constraints play a significant role, in 
particular, in our current understanding
of the possible patters of oscillations of the 
three flavour neutrinos and 
anti-neutrinos, $\nu_l$ and $\bar{\nu}_l$, 
$l=e,\mu,\tau$. The CHOOZ and Palo Verde experiments
are sensitive to values of neutrino mass 
squared difference
$\Delta m^2 \gtrsim 10^{-3}~{\rm eV^2}$,
which includes the corresponding 
atmospheric neutrino region, eq. (\ref{dmatm}).
No disappearance of the reactor  $\Pagne$
was observed. This rules out the possibility
of significant  
$\Pgngm \leftrightarrow \Pgne$ 
($\Pagngm \leftrightarrow \bar \Pagne$)
oscillations of the atmospheric
$\Pgngm$ ($\Pagngm$), and 
$\Pgne$ ($\Pagne$), which
is consistent with the 
Super-Kamiokande atmospheric neutrino 
data \cite{SKatm}. Performing a
two-neutrino oscillation
analysis, the following 
upper bound on the value of the
corresponding mixing angle, $\theta$,
was obtained by the CHOOZ collaboration
\footnote{The second possibility - of large
$\sin^2  \theta > 0.9$,
which is admitted by the CHOOZ data alone,
is incompatible with the neutrino oscillation
interpretation of the solar neutrino deficit
(see, e.g., \cite{BBGK9596,BGG99}).}
\cite{CHOOZ2} at 90\% C.L. for
$\Delta m^2 \geq 1.5 \times 10^{-3} \mathrm{eV}^2$:
\begin{equation}
\sin^{2} \theta < 0.09.
\label{chooz1}
\end{equation}
%
\noindent The precise upper limit in eq. (\ref{chooz1})
is $\Delta m^2$-dependent. 
It is a decreasing function of
$\Delta m^2$ as
$\Delta m^2$ increases up to
$\Delta m^2 \simeq 6 \cdot 10^{-3}~\mathrm{eV}^2$
with a minimum value
$\sin^{2} \theta \simeq 10^{-2}$.
The upper limit becomes an  
increasing function of
$\Delta m^2$ when
the latter increases further up to
$\Delta m^2 \simeq 8 \cdot 10^{-3}~\mathrm{eV}^2$,
where $\sin^{2} \theta < 2 \cdot 10^{-2}$ 
(see Figs. 1 - 3).
Somewhat weaker constraints on  $\sin^2\theta$
have been obtained by the Palo Verde 
collaboration \cite{PaloV}.

 The results of the two-neutrino oscillation 
analysis of the CHOOZ data remain essentially 
valid in the case of 3-neutrino mixing and oscillations,
provided the two independent neutrino mass-squared
differences characterizing the oscillations
in this case, $\Delta m^2_{21} > 0$ 
and $\Delta m^2_{31} > 0$ (see further),
obey the hierarchical relation:
\begin{equation}
 \Delta m^2_{21} \ll \Delta m^2_{31}.  
\label{hier2131}
\end{equation}
\noindent In this case one can 
 make the identification: 
\begin{equation}
\Delta m^2_{21} \simeq \Delta m^2_{\odot},~~~ 
\Delta m^2_{31}\simeq \Delta m^2_{atm} .
\label{21sol31atm}
\end{equation}
%
 Under the conditions (\ref{hier2131}) 
and (\ref{21sol31atm}) and the constraint 
(\ref{dmatm}), the oscillations of the reactor
$\Pagne$ due to  $\Delta m^2_{21}$
cannot develop on the baselines of the 
CHOOZ and Palo Verde experiments.
The corresponding 3-neutrino $\Pagne$ survival  
probability 
reduces to a two-neutrino survival
probability with 
$\Delta m^2 = \Delta m^2_{31}$
and $\sin^2\theta = |U_{e3}|^2$, 
where  $U_{e3}$ is the element of
the PMNS mixing matrix which couples the 
electron field to the heaviest neutrino field
in the mixing relation eq. (\ref{001}).
Thus, we have in the case under discussion:
\begin{equation}
\sin^2\theta = |U_{e3}|^2 = \sin^2\theta_{13},
\label{Ue3th13}
\end{equation}
%
\noindent $\theta_{13}$ being 
the one of the three mixing
angles in the standard parametrization of
the PMNS matrix (see, e.g., \cite{BGG99}).
The other two angles, $\theta_{12}$ and $\theta_{23}$,
defined through the relations
\begin{equation}
|U_{\mathrm{e} 1}| = \cos \theta_{12} \sqrt{1 - |U_{\mathrm{e} 3}|^2},
~~|U_{\mathrm{e} 2}| = \sin \theta_{12} \sqrt{1 - |U_{\mathrm{e} 
3}|^2},
\label{Ue12}
\end{equation}
\noindent and  
\begin{equation}
|U_{\mu 3}| = \sin \theta_{23} \sqrt{1 - |U_{\mathrm{e} 3}|^2},
~~|U_{\tau 3}| = \cos \theta_{23} \sqrt{1 - |U_{\mathrm{e}
3}|^2},
\label{Umutau3}
\end{equation}
\noindent control the oscillations of the solar 
$\Pgne$ and of the atmospheric 
$\Pgngm$ ($\Pagngm$), respectively.
The upper limit (\ref{chooz1}) corresponds to:
\begin{equation}
|U_{e3}|^2 = \sin^2\theta_{13} < 0.09.
\label{chooz2}
\end{equation}

\indent  Let us note that 
under the condition (\ref{hier2131}),
the solar neutrino survival probability,
$P^{3\nu}_{\odot}(\Pgne\to\Pgne)$,
which is relevant for the 
interpretation of the solar neutrino data
in terms of neutrino oscillations,
depends on $\Delta m^2_{21}$, $\theta_{12}$ and 
$|U_{e3}|^2$ (or $\theta_{13}$) (see, e.g., 
\cite{SP3nu88}): 
\begin{equation}
P^{3\nu}_{\odot}(\Pgne\to\Pgne)
\cong
\left( 1 - |U_{e3}|^2 \right)^2
P^{(2\nu)}_{\odot}(\Pgne\to\Pgne)
+
|U_{e3}|^4
\,,
\label{Pe32}
\end{equation}
%
\noindent where
$P^{(2\nu)}(\Pgne\to\Pgne)$
is the solar $\Pgne$ survival
probability in the case of two-neutrino
$\Pgne \to \nu_{\mu (\tau)}$ transitions
due to $\Delta m^2_{21}$ and $\theta_{12}$
\cite{SPexp88},
in the expression for which the solar
matter potential in the standard two-neutrino 
case, $V_{\odot} = 
\sqrt{2} G_F N_e$, $N_e$ being the solar
electron number density,
is replaced
by $(1 - |U_{e3}|^2)V_{\odot}$.
The dependence of 
$P^{(2\nu)}_{\odot}(\Pgne\to\Pgne)$ on
$\theta_{13}$, however,  is rather 
weak and cannot be used
to derive more stringent upper limit on
$\sin^2\theta_{13}$ than that obtained 
in the CHOOZ experiment, e.g., eq. (\ref{chooz2}).
Similar conclusion is valid for the upper
bounds on $|U_{e3}|^2$ which can be derived
from the Super-Kamiokande atmospheric neutrino
data (see, e.g., \cite{Kj00}).

 The element of the PMNS matrix $U_{e3}$,
more precisely, its absolute value
$|U_{e3}| =\sin\theta_{13}$, plays a very important  
role in the phenomenology of the 3-neutrino oscillations.
It drives the sub-dominant 
 $\Pgngm \leftrightarrow \Pgne$
($\Pagngm \leftrightarrow \bar \Pgne$)
oscillations of the atmospheric 
$\Pgngm$ ($\Pagngm$) and
$\Pgne$ ($\Pagne$)
\cite{SP1NOLR}. The value of 
$|U_{e3}|$ (or $\theta_{13}$)
controls also the 
$\Pgngm \rightarrow \Pgne$,
$\Pagngm \rightarrow \Pagne$,
 $\Pgne \rightarrow \Pgngm$ and
$\Pagne \rightarrow \Pagngm$
transitions in the long baseline neutrino 
oscillation experiments (MINOS, CNGS),
and in the widely discussed
very long baseline neutrino oscillation 
experiments at neutrino factories (see, e.g.,
\cite{AMMS99}). The magnitude of the 
T-violation and CP-violation effects 
in neutrino oscillations is directly proportional
to $|U_{e3}| = \sin\theta_{13}$ 
(see, e.g., \cite{KP3nu88}).
Thus, the fundamental question about 
the T- and CP-violation
in the lepton sector can be investigated 
in neutrino oscillation experiments
only if $|U_{e3}|$ is sufficiently large.
If the neutrinos with definite mass 
are Majorana particles (see, e.g., \cite{BiPet87}),
the predictions for the effective Majorana mass
parameter in neutrinoless double $\beta-$decay
depend in the case of hierarchical neutrino mass
spectrum on the value of $|U_{e3}|^2$ 
(see, e.g., \cite{BPP1}).   
The knowledge of the value of $|U_{e3}|$ 
is crucial for the searches 
for the correct theory of 
neutrino masses and mixing as well.

  In the present article we present results
of a new analysis of the CHOOZ data which limit 
$|U_{e3}|^2$. The analysis is made in the framework of the 
three-neutrino mixing hypothesis. 
The parameters characterizing 
the solar neutrino oscillations,
$\Delta m^2_{\odot}$ and 
the neutrino mixing angle
$\theta_{\odot}$, are assumed to
lie in the region of the 
LMA MSW solution of the solar 
neutrino problem. 
In the case of neutrino mass spectrum 
with normal hierarchy, for instance, 
we have, $\Delta m^2_{21}\simeq \Delta m^2_{\odot}$ and 
$\theta_{12} \simeq \theta_{\odot}$.
As we have already 
briefly discussed, the LMA MSW solution admits 
values of $\Delta m^2_{\odot}$ as large as
$\sim (5 - 8)\cdot 10^{-4}~{\rm eV^2}$ 
\cite{FogliSNO,ConchaSNO}.
For such values of  $\Delta m^2_{\odot}$
the oscillations of the reactor 
$\Pagne$ due to  $\Delta m^2_{\odot}$ 
can develop on the baselines of the CHOOZ and 
Palo Verde experiments and the contributions 
of the ``solar'' ($\Delta m^2_{\odot}$ 
and $\theta_{\odot}$)
terms in the reactor $\Pagne$
survival probability can be sizeable.
The hierarchical relation (\ref{hier2131})
is not valid in a large part of the 
interval of allowed values
of $\Delta m^2_{atm}$
(e.g., eq. (\ref{dmatm})) as well.
As a consequence, the limits on 
$|U_{e3}|^2$, obtained in
the two-neutrino oscillation analyses of
the CHOOZ and Palo Verde data 
are not valid in the case of 3-neutrino 
oscillations if $\Delta m^2_{\odot}$
lies in a rather large sub-region of the 
LMA MSW solution region. 
Using the CHOOZ data,
we obtain the 3-neutrino oscillation limits on 
$|U_{e3}|^2$ with values of
$\Delta m^2_{\odot}$ 
and $\sin^2\theta_{\odot}$
in the LMA MSW allowed region.
We find that for 
$\Delta m^2_{\odot} 
\gtrsim 2\cdot 10^{-4}~{\rm eV^2}$
 these limits are  
more stringent than the limits 
obtained from the original analysis of 
the CHOOZ data made in the framework of 
the two-neutrino mixing. 
\vskip -0.5cm
\section{Three-Neutrino Oscillations of Reactor $\Pagne$}
\vskip -0.2cm

\hskip 0.5cm  We shall number 
(without loss of generality) 
the neutrinos with definite mass in vacuum 
$\nu_j$, $j=1,2,3$, 
in such a way that their masses obey 
$m_1 < m_2 < m_3$. With this choice one has 
$\Delta m^2_{jk} > 0$ for $j > k$. We do not assume
that the hierarchy relation (\ref{hier2131})
is valid in what follows.

  Consider first the case of 
``normal hierarchy'' 
between the neutrino masses, in which 
\begin{equation}
\Delta m^2_{21} \simeq\Delta m^2_{\odot}.
\label{dm21sol}
\end{equation}
%
The neutrino mixing angle
which controls the solar 
neutrino oscillations and is 
determined in the analyses
of the solar neutrino data,
$\theta_{\odot}$, coincides in 
this case with $\theta_{12}$:
\begin{equation}
\theta_{\odot} \simeq \theta_{12}.  
\label{th12sol}
\end{equation}

\noindent For $\Delta m^2_{atm}$ 
there exist two possibilities:  
\begin{equation}
\Delta m^2_{atm} \simeq \Delta m^2_{31},  
\label{dm31atm}
\end{equation}
%
\noindent or
\begin{equation} 
\Delta m^2_{atm} \simeq \Delta m^2_{32}.
\label{dm32atm}
\end{equation}
%

  Let us note that even for  
$\Delta m^2_{\odot} \simeq \Delta m^2_{21} 
\sim (4 - 8)\cdot 10^{-4}~{\rm eV^2}$,
the corrections to 
the solar $\Pgne$
survival probability, eq. (\ref{Pe32}),
due to $\Delta m^2_{atm}$ when the 
hierarchical relation
$\Delta m^2_{\odot} \ll \Delta m^2_{atm}$
does not hold, are practically negligible
\cite{Quasiave}
and do not change the results of the analyses
of the solar neutrino data based on
the ``hierarchical'' expression 
(\ref{Pe32}) for the probability.

   The exact expression for the $\Pagne$
survival probability of interest
in the case of the three neutrino 
mixing can be written in the form
 
\begin{eqnarray}
\lefteqn{P(\Pagne\to\Pagne)} \nonumber\\
&& =\,~~ 1 - 2 \, |U_{e 3}|^2 \left( 1 - |U_{e 3}|^2 \right)
\left( 1 - \cos \frac{ \Delta{m}^2_{31} \, L }{ 2 \, E } \right)
\nonumber \\
&& - \,~~\frac{1}{2} ( 1 - |U_{e 3}|^2 )^{2}\sin ^{2}2\theta_{\odot} \,
\left( 1 - \cos \frac{ \Delta{m}^2_{21} \, L }{ 2 \, E } \right) \nonumber  \\
& & +\,~~ 2|U_{e 3}|^2 ( 1 - |U_{e 3}|^2 ) \sin^{2}\theta_{\odot}\, 
\left(\cos
\left( \frac
{\Delta{m}^2_{31} \, L }{ 2 \, E} - \frac {\Delta{m}^2_{21} \, L }{ 2 \,
E}\right)
-\cos \frac {\Delta{m}^2_{31} \, L }{ 2 \, E} \right)\, 
\label{P21sol}
\end{eqnarray}

%
\noindent In deriving the expression for 
$P(\Pagne\to\Pagne)$ we have 
utilized eqs. (\ref{Ue12}) and (\ref{th12sol}).    

  The second term in the right-hand side 
of the expression
for $P(\Pagne\to\Pagne)$, eq. (\ref{P21sol}), 
correspond to oscillations due
to $\Delta m^2_{atm}$ , the third one
is a ``solar neutrino oscillation''
term, and the fourth one is an interference 
term between the amplitudes corresponding
to the first three.
Let us notice that the 
coefficient in front of the bracket of the 
``solar neutrino oscillation'' term in eq. (\ref{P21sol}) 
is relatively large in the LMA MSW solution region, 
while the coefficient in front of the bracket
of the fourth term is of the same 
order as coefficient in front  
of the bracket in the main second term.

  In what follows we will present 
constraints on 
$|U_{e 3}|^2$ as a function of 
$\Delta m^2_{31}$ for a number of fixed values of 
$\Delta m^2_{21} \simeq \Delta m^2_{\odot}$ and 
$\sin^2\theta_{12} \simeq \sin^2\theta_{\odot}$.

   In the case of ``inverted  hierarchy'' 
between the neutrino masses one has: 
\begin{equation}
\Delta m^2_{32} \simeq \Delta m^2_{\odot}.
\label{dm32sol}
\end{equation}
%
\noindent Now $|U_{\mathrm{e} 2}|$ and
$|U_{\mathrm{e} 3}|$ are related to 
the mixing angle which controls the
solar neutrino oscillations 
$\theta_{\odot}$:
\begin{equation}
|U_{\mathrm{e} 2}| = 
\cos \theta_{\odot} \sqrt{1 - |U_{\mathrm{e} 1}|^2},
~~|U_{\mathrm{e} 3}| = 
\sin \theta_{\odot} \sqrt{1 - |U_{\mathrm{e} 
1}|^2}.
\label{Ue23}
\end{equation}
%
\noindent There are again two possibilities 
for $\Delta m^2_{atm}$: the first coincides with that
in eq. ({\ref{dm31atm}),   
while the second is given by:
\begin{equation} 
\Delta m^2_{atm} \simeq \Delta m^2_{21}.
\label{dm21atm}
\end{equation}
%
The expression
for the $\Pagne$ survival probability
of interest can be written in the following form:
\begin{eqnarray}
\lefteqn{P(\Pagne\to\Pagne)} \nonumber\\
&& =\,~~ 1 - 2 \, |U_{e 1}|^2 \left( 1 - |U_{e 1}|^2 \right)
\left( 1 - \cos \frac{ \Delta{m}^2_{31} \, L }{ 2 \, E } \right)
\nonumber \\
&& - \,~~\frac{1}{2} ( 1 - |U_{e 1}|^2 )^{2}\sin ^{2}2\theta_{\odot} \,
\left( 1 - \cos \frac{ \Delta{m}^2_{32} \, L }{ 2 \, E } \right) \nonumber  \\
& & + \,~~2|U_{e 1}|^2 ( 1 - |U_{e 1}|^2 ) \cos^{2}\theta_{\odot} \,  
\left(\cos
\left( \frac
{\Delta{m}^2_{31} \, L }{ 2 \, E} - \frac {\Delta{m}^2_{32} \, L }{ 2 \,
E}\right)
-\cos \frac {\Delta{m}^2_{31} \, L }{ 2 \, E} \right)\,, 
\label{P32sol}
\end{eqnarray}
%
\noindent where we have utilized eq. (\ref{Ue23}).

  One comment is in order. Note that, apart from the
fact that the role of $|U_{e 3}|^2$ in eq. (\ref{P21sol}) 
is played by  $|U_{e 1}|^2$ in eq. (\ref{P32sol}), 
the coefficients
in front of the last 
interference terms in eqs. (\ref{P21sol}) and (\ref{P32sol}) 
differ: in the case of eq. (\ref{P21sol})
it contains the factor
$\sin^2\theta_{\odot}$, while
the analogous factor in eq. (\ref{P32sol})
is $\cos^2\theta_{\odot}$. This implies
that for $\sin^2\theta_{\odot} = \cos^2\theta_{\odot}$,
the constraints from the CHOOZ and Palo Verde data on
$|U_{e 3}|^2$ in the case of neutrino mass spectrum
with ``normal hierarchy'' will be equivalent 
to the constraints on $|U_{e 1}|^2$
in the case of spectrum with ``inverted hierarchy''.
However, these constraints will differ if
$\sin^2\theta_{\odot} \neq \cos^2\theta_{\odot}$.
The best fit point of the solar neutrino data
in the LMA MSW solution region, for instance,
corresponds to 
$\sin^2\theta_{\odot} = 0.27$ and 
$\cos^2\theta_{\odot} = 0.73$.
Consequently,  the upper bound on 
$|U_{e 3}|^2$ in the 
``normal hierarchy'' case
can differ  
from
the upper bound on $|U_{e 1}|^2$ 
in the case of spectrum with 
``inverted hierarchy''.
\vskip -0.5cm
\section{New Constraints on $|U_{e3}|^2$ from the CHOOZ Data}
\vskip -0.2cm

\hskip 0.5cm  Taking into account the possibility of 
relatively large values of 
$\Delta{m}^2_{\odot}$, which are allowed in the case
of the  most favorable by the data 
LMA solution of solar neutrino
problem, we have re-analyzed the CHOOZ data
and have obtained 
exclusion curves that took into 
account the contribution of the
``solar terms'' in the reactor $\Pagne$
survival probability (eq. (\ref{P21sol}) and eq. (\ref{P32sol})). 
We will present results derived 
for two values of the 
solar neutrino mixing angle, 
$\sin ^{2}\theta_{\odot} = 0.5;~0.27$,
and for 
$\Delta m_{\odot}^{2} = 0;~2\cdot 10^{-4};~
4\cdot 10^{-4};~6\cdot 10^{-4}~{\rm eV^2}$.
All these values of 
$\sin ^{2}\theta_{\odot}$
and $\Delta m_{\odot}^{2}$
lie in the region of the 
LMA MSW solution of the solar neutrino problem
\cite{FogliSNO,ConchaSNO,GoswaSNO,StrumiaSNO,GiuntiSNO}.
\vskip -0.5 cm
\subsection{The CHOOZ Data and their Statistical Analysis}
\vskip -0.2cm

\hskip 0.5truecm The CHOOZ experiment, 
just like the other 
reactor experiments, detected the 
$\Pagne$'s produced by each reactor 
of the homonym power plant through 
the reaction
\begin{equation}
\Pagne + \Pp \rightarrow \Pn + \Pep
\label{eq:betainv}
\end{equation}
%
The $\Pep$ energy deposit in the 
detector is strongly correlated with the incoming 
$\Pagne$ energy (the correlation being 
linear as long as the proton recoil can
be neglected). An evidence for neutrino 
oscillations in CHOOZ could thus 
result from:
\begin{enumerate} 
\item a deficit of the $\Pagne$ detection 
rate with respect to expectations; 
\item distortion in the ratio of the 
measured positron spectrum versus 
predictions.
\end{enumerate}

  The most powerful test (analysis ``A''\cite{CHOOZ2}) 
combine both pieces of 
information. The event sample is divided 
into seven $\Pep$ energy bins ranging
from $0.8$ to $6.4 \,\rm{MeV}$; for each energy bin, 
the ramp-up and down of 
reactors during the data taking period 
allowed to subtract the
background as well as to extract the 
contribution of each reactor to the signal
rate \cite{mythesis}. As a result, we have two 
$\Pep$ experimental spectra at 
our disposal, to be compared with the predictions. 
The expected positron 
spectrum is obtained by Monte Carlo simulation 
of the detector response, after
folding the no-oscillation $\Pagne$ 
flux with the neutrino survival
probability (eq. (\ref{P21sol}) or (\ref{P32sol})
depending on whether normal or 
inverted hierarchy is considered); 
so the positron yield for the k-th reactor
and the j-th energy bin can be parametrized as follows:
\begin{equation}
  \overline{Y}(E_j,L_k,\theta,\Delta m^2_{31}) = \tilde{Y}(E_j) 
  \overline{P}(E_j,L_k,|U_{e3}|^2,\Delta m^2_{31}),
  \quad (j=1,...,7,~k=1,2)
  \label{xosc}
\end{equation}
%
where $\tilde{Y}(E_j)$ is the distance-independent 
positron yield in the absence
of neutrino oscillations, $L_k$ is the 
reactor-detector distance and the 
last factor represents the survival 
probability averaged over the energy 
bin and the finite detector and reactor 
core size \cite{mythesis}. Both measured 
and predicted yields can then be 
arranged into 14-element arrays and, 
in order to test the compatibility of a 
certain oscillation hypothesis $(|U_{e3}|^2,\Delta m^2_{31}$) 
with measurements, we 
can build the following $\chi^2$ 
function
\begin{multline}
  \chi^2 \bigl(|U_{e3}|^2,\Delta m^2_{31},\alpha,g \bigr) = \\
   \sum_{i,j=1}^{14}
  \Bigl( Y_i - \alpha \overline{Y} \bigl(gE_i,L_i,|U_{e3}|^2,\delta m^2_{31} 
\bigr)
  \Bigr) V_{ij}^{-1}
  \Bigl( Y_j - \alpha \overline{Y} \bigl(gE_j,L_j,|U_{e3}|^2,\delta m^2_{31} 
\bigr)
  \Bigr) + \\
  \left( \frac{\alpha-1}{\sigma_\alpha} \right)^2 + \left( \frac{g-1}{\sigma_g}
  \right)^2,
  \label{chi2}
\end{multline}
%
\noindent where $V_{ij}$ is the error matrix, $\alpha$ is the absolute 
normalization constant 
($\sigma_\alpha = 2.7\%$), $g$ is the 
energy-scale calibration factor ($\sigma_g=1.1\%$), 
$L_i= L_1 \text{ for } i\leq 7$ and $L_i= L_2 \text{ for } i> 7$.
The $\chi^2$ value for a certain 
parameter set is obtained by minimizing
(\ref{chi2}) with respect to $\alpha$ and $g$.

   Confidence intervals at $90\%$ C.L. 
can be obtained by using a frequentist 
approach, according to the 
Feldman and Cousins prescription \cite{FeldCou}. 
The ``ordering'' principle is 
based on the logarithm of the ratio 
of the likelihood functions for the two cases:
\begin{equation}
  \lambda(|U_{e3}|^2,\Delta m^2_{31}) = \chi^2(|U_{e3}|^2,\Delta m^2_{31}) - 
\chi^2_{min}
  \label{loglikrat}
\end{equation}
%
where the minimum $\chi^2$ value 
must be searched for within the physical 
domain ($0<|U_{e3}|<1$, $\Delta m^2_{31}>0$).
Smaller $\lambda$ values imply a better 
agreement of the hypothesis with the 
data. The $\lambda$ distribution for 
the given parameter set was evaluated by 
performing a Monte Carlo simulation of 
a large number (5000) of experimental 
positron spectra whose values are scattered 
around the predicted positron yields
$\overline{Y}(E_i,L_i,|U_{e3}|^2,\Delta m^2_{31})$ 
with Gaussian-assumed variances. For each set
we extracted the quantity 
$\lambda_c(|U_{e3}|^2,\Delta m^2_{31})$ 
such that $90\%$ of the
simulated experiments have 
$\lambda<\lambda_c$. The $90\%$ 
confidence domain 
then includes all points in 
the $(|U_{e3}|^2,\Delta m^2_{31})$ plane such that
\begin{equation}
  \lambda_{exp}(|U_{e3}|^2,\Delta m^2_{31}) < 
\lambda_c(|U_{e3}|^2,\Delta m^2_{31}),
  \label{clord}
\end{equation}
where $\lambda_{exp}$ is evaluated for 
the experimental data for each point in 
the physical domain. 

  The above procedure is then repeated 
to extract confidence intervals for 
different values of the solar neutrino 
oscillation parameters entering 
eqs. (\ref{P21sol}) and (\ref{P32sol}).
The results are shown in 
Fig.~\ref{fig:norm05} (normal hierarchy, 
$\sin^2\theta_\odot = 0.5$), 
Fig.~\ref{fig:norm027} (same as before, 
with $\sin^2\theta_\odot = 0.27$) and
Fig.~\ref{fig:inv} (inverted hierarchy, $\sin^2\theta_\odot = 0.27$).
All parameters lying to the right of the contours are 
excluded by the CHOOZ data at 90\% C.L., while the parameter
regions to the left of the contours
are compatible with the data.
\vskip -0.5cm
\subsection{The Results}
\vskip -0.2cm
\hskip 0.5 truecm
As Figs. 1 - 3 show, the new 
exclusion curves are practically
indistinguishable from the curves 
found in  the two-neutrino oscillation analysis 
of the CHOOZ data if  $\Delta m_{\odot}^{2}
\lesssim 10^{-4}\rm{eV}^{2}$.
Thus, the CHOOZ bound on $|U_{e3}|^{2}$ is stable 
with respect to corrections due to the ``solar''
terms as long as $\Delta m_{\odot}^{2}$ does not exceed
$\sim 10^{-4}\rm{eV}^{2}$.

  The 3-neutrino oscillation analysis leads, 
however, to more stringent constraints 
than the two-neutrino
oscillation analysis if
$\Delta m_{\odot}^{2}
\gtrsim 2\times 10^{-4}~\rm{eV}^{2}$, as 
it is seen in Figs. 1 - 3. For 
$\Delta m_{\odot}^{2}
\cong (4 - 6)\cdot 10^{-4}~\rm{eV}^{2}$
we find that the upper bounds on 
$|U_{e3}|^{2}$ ($|U_{e1}|^{2}$) are 
considerably more stringent 
than those derived in the two-neutrino
oscillation analysis of the CHOOZ data.
The upper bounds of interest depend on
$\sin^2\theta_{\odot}$.
At $\Delta m^{2}_{31} =
2.5\cdot 10^{-2}~\rm{eV}^2$, which is the best 
fit value of the Super-Kamiokande 
atmospheric neutrino data, and 
for $\Delta m^{2}_{\odot} = 4.0\cdot10^{-4}~\rm{eV}^2$,
we get for $\sin ^{2}\theta_{\odot} = 0.50~(0.27)$:
\begin{equation}
|U_{e3}|^{2}\leq 2.9\cdot 10^{-2}~ (3.0\cdot 10^{-2}). 
\end{equation}

\noindent For $\Delta m^{2}_{\odot} = 6.0 \cdot10^{-4}~\rm{eV}^2$,
and $\sin ^{2}\theta_{\odot} = 0.50$ the same limit reads:
\begin{equation}
|U_{e3}|^{2}\leq 1.7\cdot 10^{-2},
\end{equation}
%
\noindent while if $\sin ^{2}\theta_{\odot} = 0.27$, we get
$|U_{e3}|^{2}\leq 2.0\cdot 10^{-2}$.
The limits on $|U_{e3}|^{2}$ ($|U_{e1}|^{2}$ 
in the case of the inverted 
hierarchy) for different values of solar 
neutrino oscillation parameters and different
values of $\Delta m^{2}_{31}$ are presented 
in  Table 1.

\begin{table}[htbp]
  \caption{\small Limits on the $\Pgne$ mixing parameter $|U_{e3}|^2$ 
($|U_{e1}|^2$ in the case of inverted hierarchy) for three values of 
$\Delta m^2_{31}$ and for different values of solar neutrino 
oscillation parameters.}
  \label{tab:mixlim}
  \begin{center}
    \begin{tabular}{|c|c|c|c|c|}
      \hline
        $\Delta m^2_{31}$ & $\Delta m^2_\odot$ &
        $|U_{e3}|^2$ & $|U_{e3}|^2$ & $|U_{e1}|^2$ \\
        $({\rm eV}^2)$ & $({\rm eV}^2)$ &  
        $(\sin^2 \theta_\odot = 0.5)$ & $(\sin^2 \theta_\odot = 0.27)$ &
        $(\sin^2 \theta_\odot = 0.27)$ \\
      \hline
      \hline
        $2.5\cdot 10^{-3}$ & $0$ & 
        \multicolumn{3}{c}{$3.7\cdot 10^{-2}$} \vline \\
      \cline{2-5}
        & $2\cdot 10^{-4}$ & $3.6\cdot 10^{-2}$ & $3.6\cdot 10^{-2}$ 
        & $3.8\cdot 10^{-2}$ \\
      \cline{2-5}
        & $4\cdot 10^{-4}$ & $2.9\cdot 10^{-2}$ & $3.0\cdot 10^{-2}$ 
        & $3.5\cdot 10^{-2}$ \\
      \cline{2-5}
        & $6\cdot 10^{-4}$ & $1.7\cdot 10^{-2}$ & $2.0\cdot 10^{-2}$ 
        & $2.6\cdot 10^{-2}$ \\
      \hline
      \hline
        $10^{-2}$ & $0$ & 
        \multicolumn{3}{c}{$3.6\cdot 10^{-2}$} \vline \\
      \cline{2-5}
        & $2\cdot 10^{-4}$ & $3.4\cdot 10^{-2}$ & $3.4\cdot 10^{-2}$ 
        & $3.4\cdot 10^{-2}$ \\
      \cline{2-5}
        & $4\cdot 10^{-4}$ & $2.8\cdot 10^{-2}$ & $2.9\cdot 10^{-2}$ 
        & $2.9\cdot 10^{-2}$ \\
      \cline{2-5}
        & $6\cdot 10^{-4}$ & $1.7\cdot 10^{-2}$ & $2.1\cdot 10^{-2}$ 
        & $2.0\cdot 10^{-2}$ \\
      \hline
      \hline
        $10^{-1}$ & $0$ & 
        \multicolumn{3}{c}{$2.5\cdot 10^{-2}$} \vline \\
      \cline{2-5}
        & $2\cdot 10^{-4}$ & $2.3\cdot 10^{-2}$ & $2.3\cdot 10^{-2}$ 
        & $2.3\cdot 10^{-2}$ \\
      \cline{2-5}
        & $4\cdot 10^{-4}$ & $2.0\cdot 10^{-2}$ & $2.1\cdot 10^{-2}$ 
        & $2.1\cdot 10^{-2}$ \\
      \cline{2-5}
        & $6\cdot 10^{-4}$ & $1.2\cdot 10^{-2}$ & $1.6\cdot 10^{-2}$ 
        & $1.6\cdot 10^{-2}$ \\
      \hline
      \hline
    \end{tabular}
  \end{center}
\end{table}

 The results from the ongoing experiments with solar neutrinos
 Super-Kamiokande, SNO, SAGE and GNO, as well as from the future
experiment BOREXINO \cite{BOREX}, will lead to a considerable progress
in the searches for the correct and unique
solution of the solar neutrino problem. 
The LMA MSW solution will be tested in 
detail by the KamLAND
experiment \cite{KamL}. The results of these 
experiments will allow to make a more definite 
conclusion concerning the possibility of 
large ``solar term'' effects on the limits of
the important parameter $|U_{e3}|^{2}$, that can be inferred from 
the data of the CHOOZ 
and Palo Verde experiments.
\vskip -0.5cm
\section{Conclusions}
\vskip -0.2cm
  
   In conclusion, we have presented here the results
of a new three-neutrino oscillation analysis of the CHOOZ data, 
which imposes stringent constraints on the oscillations
of electron (anti-)neutrinos.
The earlier obtained CHOOZ constraints 
are valid in the case of two-neutrino 
oscillations, as well as in the case of 
3-neutrino oscillations if 
the neutrino mass-squared differences
characterizing the oscillations of the
solar and atmospheric neutrinos,
$\Delta m^2_{\odot}$
and $\Delta m^2_{atm}$,
obey the hierarchical relation
$\Delta m^2_{\odot} \ll \Delta m^2_{atm}$.
In such a way, stringent upper 
limits on the $|U_{e3}|^{2}$ element of the
PMNS matrix
have been obtained.
 
  In our analysis the parameters 
characterizing 
the solar neutrino oscillations,
$\Delta m^2_{\odot}$ and 
the neutrino mixing angle
$\theta_{\odot}$, were assumed to
lie in the region of the 
LMA MSW solution of the
solar neutrino problem,
which is favored by the current 
solar neutrino data, 
including the SNO results.
This solutions admits values of 
$\Delta m^2_{\odot}$ as large as
$\sim (5 - 8)\cdot 10^{-4}~{\rm eV^2}$. 
For $\Delta m^2_{\odot} 
\gtrsim 2\cdot 10^{-4}~{\rm eV^2}$
the oscillations of the rector 
$\Pagne$ due to  $\Delta m^2_{\odot}$ 
can develop on the baselines of the CHOOZ and 
Palo Verde experiments and the contributions 
of the ``solar''  terms in the reactor $\Pagne$
survival probability can be sizeable.
In addition, for the
indicated values of
$\Delta m^2_{\odot}$
the hierarchical relation between
the ``solar''  $\Delta m^2_{\odot}$
and the ``atmospheric'' 
$\Delta m^2_{atm}$,
does not hold
in a large part of the 
interval of allowed values
of $\Delta m^2_{atm}$.
As a consequence, the limits on 
$|U_{e3}|^2$, obtained in
the previous neutrino
oscillation analyses of
the CHOOZ and Palo Verde data 
are not valid 
for $\Delta m^2_{\odot}$
lying in a rather large sub-region of the 
LMA MSW solution region.
We have obtained upper limits 
on $|U_{e3}|^2$ in the 
indicated sub-region of 
values of $\Delta m^2_{\odot}$.

  We have considered also the case of 
``inverted hierarchy'' between neutrino masses,
which differs under the conditions of interest
from the case of ``normal hierarchy''.
Limits on the corresponding element
of the PMNS matrix,
$|U_{e 1}|^2$, 
in the case of 
neutrino mass spectrum with
inverted mass hierarchy 
were obtained as well. 

 Our analysis was
based on the comparison of the positron spectra
measured in the CHOOZ experiment with that 
expected in the case of
3-neutrino neutrino oscillations.
We have found, in particular, that
at $\Delta m^{2}_{\odot} \lesssim 
10^{-4} \rm{eV}^{2}$, the
effect of the ``solar''
$\Delta m^{2}_{\odot}$ and $\theta_{\odot}$ 
on the $\Pagne$ survival probability 
is very small and the CHOOZ 
upper bound on $|U_{e3}|^2$
practically coincides with 
that obtained in 
the original two-neutrino mixing analysis.
For $\Delta m^{2}_{\odot} \gtrsim 2\cdot
10^{-4}~{\rm eV^2}$, the constraints
of the CHOOZ data on $|U_{e3}|^2$
are more stringent 
than in the case of 
$\Delta m^{2}_{\odot} \lesssim 
10^{-4} \rm{eV}^{2}$. 
They depend on the value of
$\sin^2\theta_{\odot}$.
For, e.g., $\sin^2\theta_{\odot} = 0.50$
and $\Delta m^{2}_{\odot} \simeq 4\cdot
10^{-4}~(6\cdot 10^{-4}) \rm{eV}^{2}$, 
we find that at 
$\Delta m^{2}_{31} = 2.5\cdot 10^{-3}~{\rm eV^2}$
(the best fit-point of the Super-Kamiokande 
atmospheric neutrino
data), one has $|U_{e3}|^{2} < 2.9 \cdot 10^{-2}$
($|U_{e3}|^{2} < 1.7 \cdot 10^{-2}$),
while for ``large'' $\Delta m^{2}_{31}$ 
we found $|U_{e3}|^{2} < 2 \cdot 10^{-2}$ 
( $|U_{e3}|^{2} < 1.2 \cdot 10^{-2}$).
These upper bounds are approximately by a factor of 
1.3 (2.2)  smaller then the upper
bound obtained in the two-neutrino mixing 
analysis of the CHOOZ data.
Let us stress that the value of the parameter
$|U_{e3}|^{2}$ ($|U_{e1}|^{2}$ ) 
is extremely important for the
future Super Beam and Neutrino 
Factory programs. In particular, 
the possibility
to investigate CP violation in the 
lepton sector at these facilities 
depends on the value of this parameter.

 We expect that a three-neutrino oscillation
anlysis of the Palo Verde data, performed under the 
conditions of the analysis
made in the present article, 
will lead at $\Delta m^{2}_{\odot} \gtrsim 2\cdot
10^{-4}~{\rm eV^2}$
to more stringent constraints on 
$|U_{e3}|^{2}$ (or $|U_{e1}|^{2}$)
than the two-neutrino oscillation analysis
of the Palo Verde data \cite{PaloV}.

\vskip 0.2truecm
\noindent {\bf Acknowledgements} 
We would like to thank the CHOOZ collaboration for 
many fruitful discussions and support.
S.M.B. would like to thank the Alexander von Humboldt Foundation 
for support. S.T.P. acknowledges with gratefulness
the hospitality and support of the SLAC Theoretical 
Physics Group, where part of the work on 
the present study was done.

\begin{figure}[hp]
  \begin{center}
    \mbox{\includegraphics[width=\linewidth]{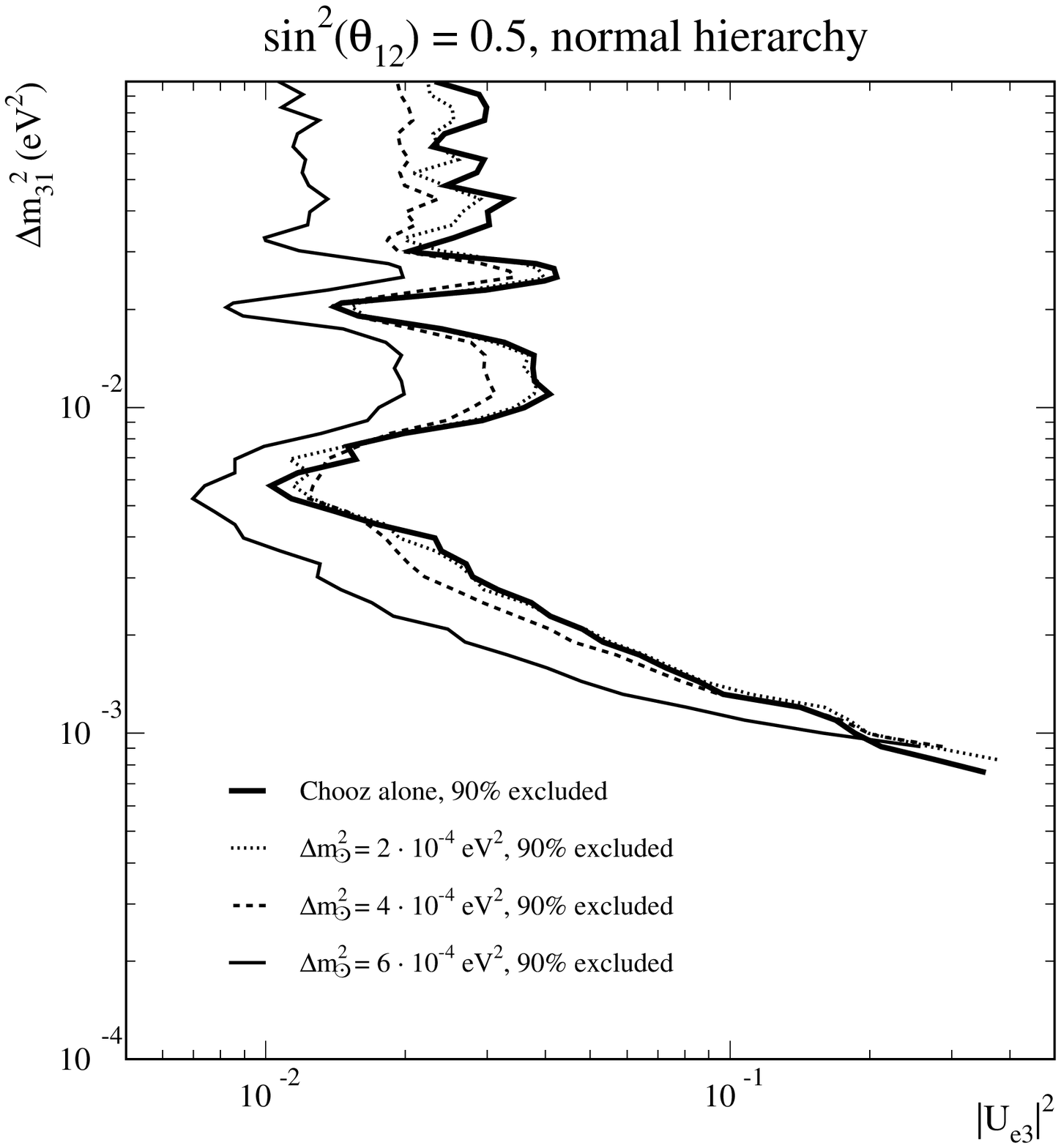}}
    \caption{\small Exclusion plot (90\% C.L.) obtained 
in the case of normal neutrino mass hierarchy for 
maximum ``solar'' mixing angle 
$\sin^2_{\odot} = 0.5$ ($\theta_{12} = \theta_{\odot}$), 
and for values of $\Delta m^2_\odot$ from the
LMA solution region. The CHOOZ result obtained 
in the case of two-neutrino mixing is 
also shown (doubly thick solid line).}
    \label{fig:norm05}
  \end{center}
\end{figure}
\begin{figure}[htbp]
  \begin{center}
    \mbox{\includegraphics[width=\linewidth]{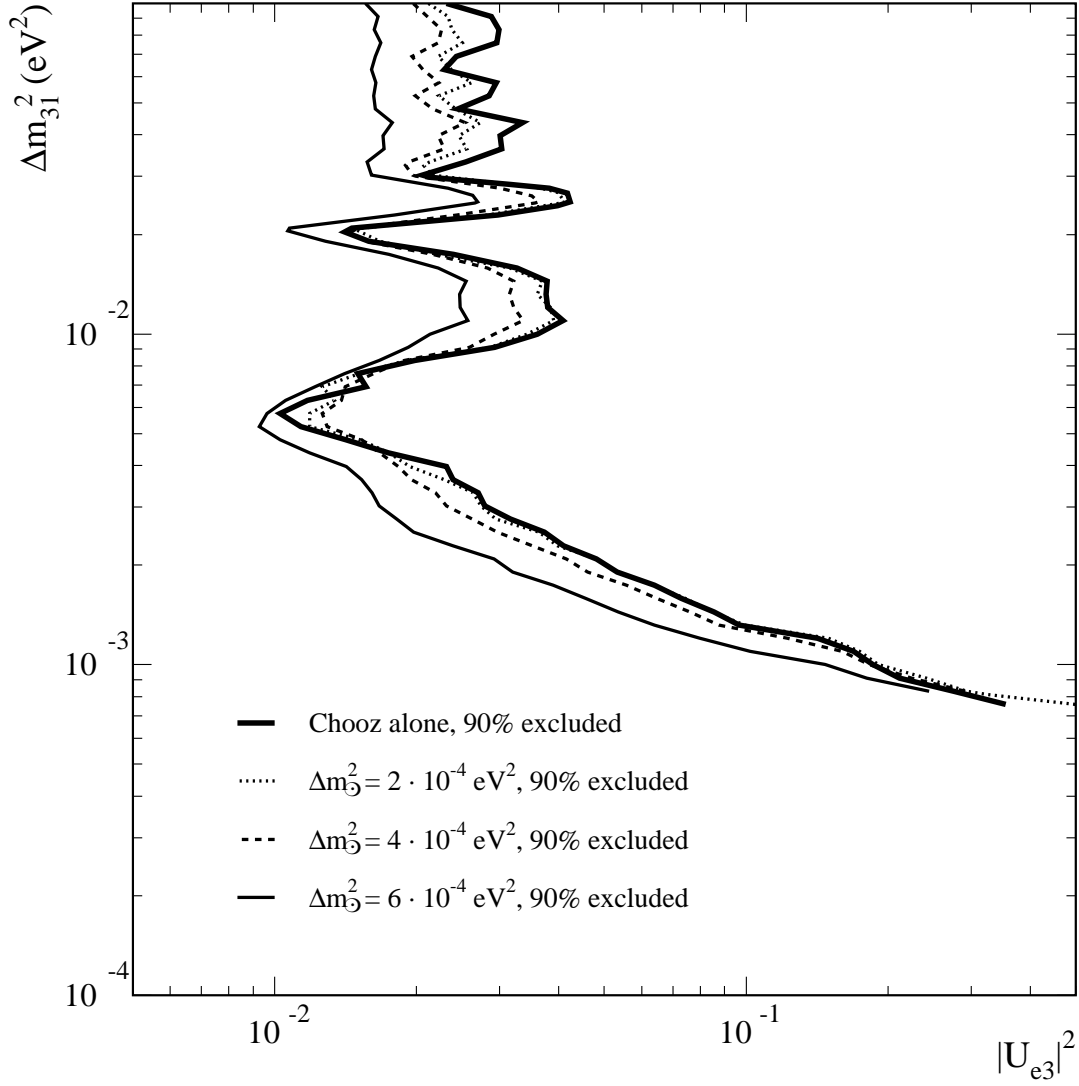}}
    \caption{\small The same as in Fig. 1 for
$\sin^2\theta_\odot = 0.27$, 
corresponding to the best fit value of the
``solar'' mixing angle ($\theta_{\odot} = \theta_{12}$) 
in the LMA region.}
    \label{fig:norm027}
  \end{center}
\end{figure}
\begin{figure}[htbp]
  \begin{center}
    \mbox{\includegraphics[width=\linewidth]{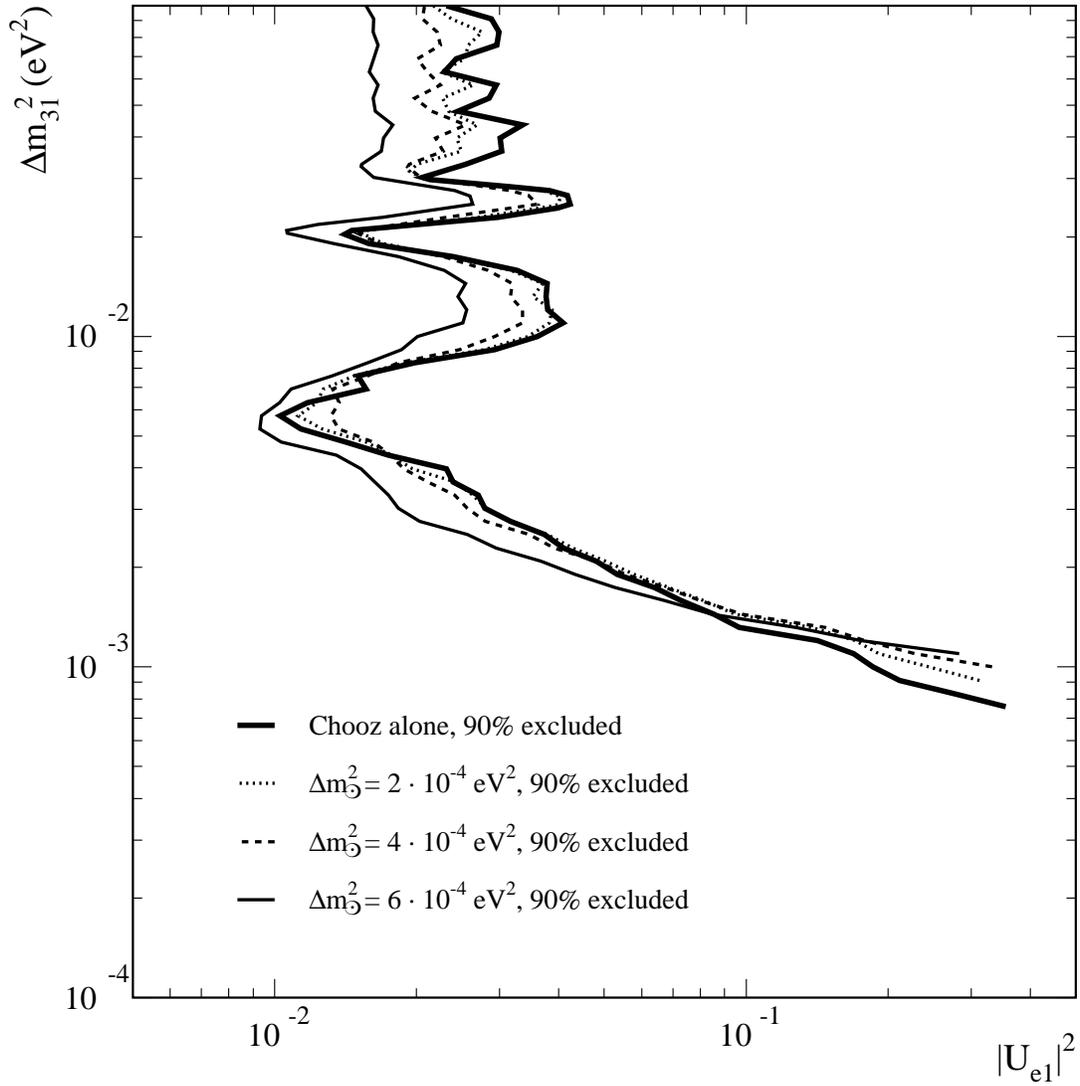}}
    \caption{\small The same as in Fig. 2 
in the case of inverted hierarchy.}
    \label{fig:inv}
  \end{center}
\end{figure}

\end{document}